\documentclass[lettersize,journal]{IEEEtran}
\usepackage{amsmath,amsfonts}
\usepackage{array}
\usepackage[caption=false,font=normalsize,labelfont=sf,textfont=sf]{subfig}
\usepackage{textcomp}
\usepackage{stfloats}
\usepackage{url}
\usepackage{verbatim}
\usepackage{graphicx}
\usepackage{soul}
\soulregister{\cite}7
\soulregister{\ref}7
\soulregister{\textit}7
\soulregister{\circled}7
\usepackage{tabularx}
\usepackage{color, xcolor}

\def\BibTeX{{\rm B\kern-.05em{\sc i\kern-.025em b}\kern-.08em
    T\kern-.1667em\lower.7ex\hbox{E}\kern-.125emX}}
\usepackage{balance}

\newcommand{\pluseq}{\mathrel{+}=}
\newcommand{\ie}{\textit{i.e.}}
\newcommand{\eg}{\textit{e.g.}}

\usepackage[ruled]{algorithm2e}
\usepackage{algpseudocode}

\usepackage{tikz}
\newcommand*\circled[1]{\tikz[baseline=(char.base)]{
            \node[shape=circle,draw,inner sep=0.7pt] (char) {#1};}}

\usepackage{graphicx}
\usepackage{multirow}
\usepackage{booktabs}
\usepackage{makecell}

\begin{document}

\title{Variation Enhanced Attacks Against RRAM-based Neuromorphic Computing System}

\author{\IEEEauthorblockN{
Hao Lv$^{1,2,}$\IEEEauthorrefmark{1},
Bing Li$^{4,}$\IEEEauthorrefmark{1}, 
Lei Zhang$^{1}$, 
Cheng Liu$^{1}$, 
Ying Wang$^{1,2,3,}$\IEEEauthorrefmark{2}}, \\
\IEEEauthorblockA{
\textit{Institute of Computing Technology, Chinese Academy of Sciences$^1$,}\\
\textit{University of Chinese Academy of Sciences$^2$,}\\
\textit{State Key Laboratory of Computer Architecture$^3$},
\textit{Capital Normal University$^4$}\\
\textit{Email}: {\textit{\{lvhao21b, zlei,liucheng, wangying2009\}@ict.ac.cn},
\textit{bing.li@cnu.edu.cn}
}}
\thanks{
Copyright © 2022, IEEE. Personal use of this material is permitted.
However, permission to use this material for any other purposes must be
obtained from the IEEE by sending an email to pubs-permissions@ieee.org.

TCAD Sep. 2022, DOI: 10.1109/TCAD.2022.3207316

https://ieeexplore.ieee.org/document/9893887

\IEEEauthorrefmark{1}Both authors contributed equally to this work.

\IEEEauthorrefmark{2}Corresponding author.}
}

\markboth{
}%
{
}

\maketitle

\begin{abstract}

The RRAM-based neuromorphic computing system (NCS) has amassed explosive interests for its superior data processing capability and energy efficiency than traditional architectures, and thus being widely used in many data-centric applications. The reliability and security issues of the NCS therefore become an essential problem. In this paper, we systematically investigated the adversarial threats to the RRAM-based NCS and observed that the RRAM hardware feature can be leveraged to strengthen the attack effect, which hasn't been granted sufficient attention by previous algorithmic attack methods.
Thus, we proposed two types of hardware-aware attack methods with respect to different attack scenarios and objectives. The first is adversarial attack, VADER, which perturbs the input samples to mislead the prediction of neural networks. The second is fault injection attack, EFI, which perturbs the network parameter space such that a specified sample will be classified to a target label, while maintaining the prediction accuracy on other samples. Both attack methods leverage the RRAM properties to improve the performance compared with the conventional attack methods.
Experimental results show that our hardware-aware attack methods can achieve nearly 100\% attack success rate with extremely low operational cost, while maintaining the attack stealthiness.

\end{abstract}

\section{Introduction}
\IEEEPARstart{T}{he} neuromorphic computing system (NCS) has attracted extensive interests as the traditional Von-Neumann architecture based on the CMOS technology is approaching the physical limit and facing the challenges of the well-known ``memory wall''.
Recent advancements in the neuromorphic algorithms (\ie, deep neural networks, DNNs) have achieved tremendous success in the computing vision (\cite{he2016deep, simonyan2014very, ren2015faster}) and natural language processing domains (\cite{devlin2018bert, sutskever2014sequence}), driving the development of the NCS towards the ultimate goal of emulating the biologic neural networks. 
Unfortunately, the neuromorphic computing system based on the traditional digital hardware suffers the degraded performance and power efficiency because it is impractical to store the synaptic weights of the deep neural models in the on-chip memory.

Therefore, the emerging non-volatile memory (eNVM) to implement the NCS has received tremendous attention due to the merits of representing the synaptic weight with the cell resistance instead of the electronic charges (\cite{jo2010nanoscale, indiveri2013integration}).
Amongst the candidate eNVM technologies, the Resistive Random Access Memory (RRAM), a.k.a. memristor, is one of the most promising ones as the natural similarity between the programmable resistance and the variable biological synaptic strengths.  
More importantly, an RRAM crossbar is a natural dot-product engine that can process the most essential operations for the neural networks in a highly efficient way (\cite{xia2016switched}). 
Recent works have reported system-level RRAM-based NCS platforms~(\cite{shafiee2016isaac, chi2016prime, song2017pipelayer,li2020hitm,li2020red}), demonstrating the powerful capability and flexibility.

On the other hand, the security of the NCS raises a concern as the DNN models are spreading into the safety-sensitive scenarios, and previous studies have demonstrated that the DNN models are vulnerable to the well-designed attacks (\cite{goodfellow2014explaining, madry2017towards, liu2017fault, zhao2019fault}).
To fool the neural networks, these attack methods either perturb the input samples (\ie, adversarial attack~\cite{madry2017towards, szegedy2013intriguing}) or manipulate the network parameters (\ie, fault injection \cite{liu2017fault, zhao2019fault}). 
By distorting the selected input or network parameters, the attackers can mislead the network to make an erroneous prediction.

We observed that the inherent variation of RRAM-based NCS could be a potential hardware-level threat to the security of the neural network (detailed in section \ref{security_risk}.), which motivates us to design hardware-aware attacks that incorporate the hardware information to improve the attack performance.
As a type of nanoscale device, RRAMs suffer from nonideal properties such as resistance fluctuation, resistance drift, and random noise~(\cite{kannan2015modeling, he2019noise}).
Though prior works~(\cite{he2019noise,liu2017rescuing,li2019build}) have been proposed to mitigate the negative effect of these hardware variations to ensure the reliability of the  RRAM-based NCS, it will be a different story when considering the security of RRAM-based NCS in the presence of the RRAM variation.

In this work, we develop two attack schemes for the RRAM-based NCS by leveraging the RRAM variation issue and the programmable cell resistance corresponding to the synapse weight. To the best of our knowledge, this is the first work that comprehensively investigates the security of the RRAM-based NCS.

Our contribution can be summarized as follows:
\begin{itemize}
\item We observed that the intrinsic variations in the RRAM pose a potential threat for the security of RRAM-based NCS, and proposed two powerful attack methods, VADER and EFI, by exploiting the hardware properties of the RRAM. In specific, the VADER focuses on poisoning the input samples while EFI targets distorting the DNN parameters to launch the effective attack on the RRAM-based NCS. 

\item In the VADER, we propose a \textit{variation amplification} algorithm to locate and perturb the variation-sensitive pixels in the input sample, which can maximize the impact of the RRAM variation on the computing results of the RRAM-based NCS. In the end, VADER effectively misleads RRAM-based NCS and makes it output the prediction error.

\item 
In the EFI, we propose a \textit{greedy victim parameter selection} algorithm to select and distort the victim parameters. By exploiting the RRAM variation, EFI can minimize the required victim parameters for a successful fault injection attack, significantly saving the operational cost.

\item 
We perform comprehensive experiments to evaluate the stealthiness, effectiveness and efficiency of our proposed attack methods on the RRAM-based NCS and compare our work with the classical attack methods~(\cite{liu2017fault, zhao2019fault, madry2017towards}).
The results show VADER and EFI are more effective than the comparison counterparts and both achieve almost 100$\%$ attack success rate. Besides, EFI can save orders of the cost (\textgreater 95\%) for the fault injection, while maintaining less accuracy degradation compared with previous fault injection attacks (\cite{liu2017fault, zhao2019fault}).
\end{itemize}

The rest of the paper is organized as follows. In section II, we introduce the background of the RRAM-based computing system, an overview of RRAM variation, and discuss the security concerns of the computing system. Section III and Section IV present our proposed attack methods for different attack scenarios in detail. Experimental results and discussion of the defense techniques for the proposed attacks are provided in Section V. Finally, Section VI concludes the paper.
\section{Background and Related Work}

\subsection{RRAM-based Neuromorphic Computing System}

\begin{figure*}[t]
  \centering
  \includegraphics[width=0.95\linewidth]{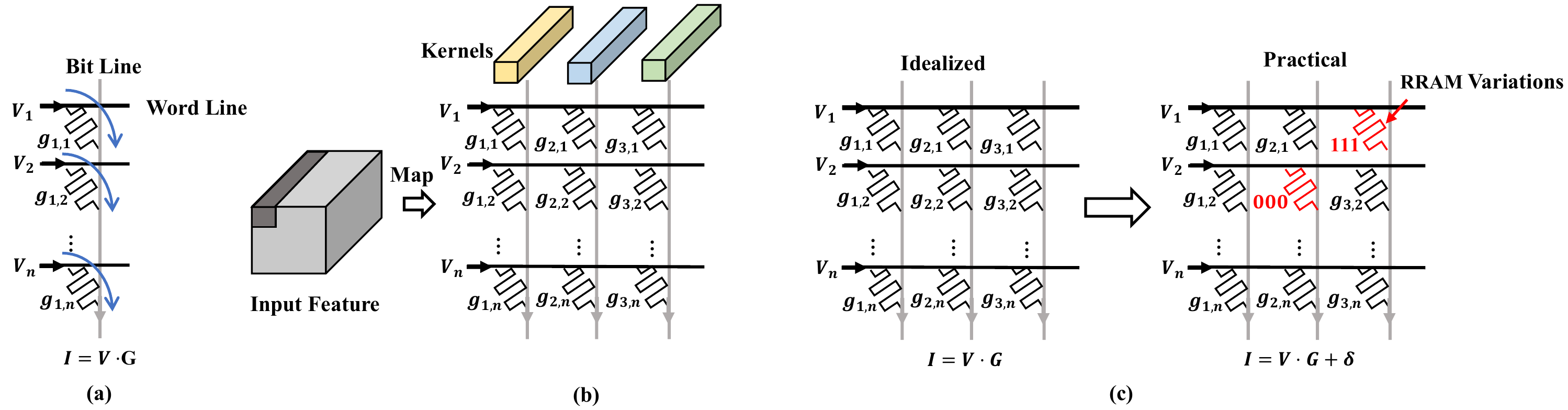}
  \caption{(a)RRAM-based analog dot-product computing; (b) An example for accelerating the neural networks. (c) An simple example for the RRAM variations.}
  \label{rram_basics}
  \vspace{0.5\baselineskip}
\end{figure*}
Features by the non-volatility and programmable resistance of the device, the RRAM has been studied for establishing the neuromorphic computing systems (\cite{jo2010nanoscale, indiveri2013integration}).
Besides, the RRAM-based NCS relies on the crossbar structure to mimic the heart operation, \ie, matrix-vector multiplication (MVM), in neural networks, which is expressed as $I=V\cdot G$ in the analog manner.
As illustrated in Figure~\ref{rram_basics}(a), the input voltage vector is applied on the word line (WL), the conductance values of the RRAM cells $G$ representing the weight matrix. The current $I$ through each bit line (BL) is naturally the multiplication result $V\cdot G$ according to the Kirchhoff's and Ohm's law~(\cite{hu2016dot}).
In this way, the RRAM crossbar structure realizes MVM calculation with a time complexity of only $O(1)$.
Figure~\ref{rram_basics}(b) shows an exemplary implementation of convolutional NNs (CNNs) on the RRAM crossbar by programming the convolutional kernels on the BLs of the crossbar and unfolding the feature maps into the input vector.
Extensive researches have explored various RRAM-based NCS designs for a variety of neuromorphic computing models applied in different application fields~(\cite{shafiee2016isaac, chi2016prime, song2017pipelayer, li2020red}).

\subsection{RRAM Variation} 
As an emerging nanoscale device, the RRAM suffers from a variety of variation issues such as manufacturing defect, resistance variation and random noise due to the immature manufacture and the intrinsic nature~(\cite{he2019noise}).
For example, the defect cells in an RRAM crossbar stuck at a certain resistance state and are not changeable~(\cite{li2019build}), while some RRAM devices' readout resistance is slightly biased from the programmed one affected by the resistance variation~(\cite{he2019noise}). 
Therefore, the network model mapped in the practical RRAM-based NCS is different from the algorithmic model, and will induce subtle computational error as Figure~\ref{rram_basics}(c) shows. 
For simplicity, we denote the cells with significant variation issues as faults in this work.
Extensive studies have investigated the effect of RRAM variation issues on the reliability of RRAM-based NCS and proposed various strategies to alleviate this problem.
\cite{kannan2015modeling} developed an efficient test scheme for detecting and locating the position of faults in the RRAM array.
With the fault distribution of the RRAM array, \cite{xia2017fault} proposed the fault-resilient training strategy to rescue the accuracy of the neural network models by exploiting the intrinsic error resilience of the neural network. 
\cite{liu2017rescuing} remapped the significant weight neural network model to the RRAM-based architecture by introducing the redundant hardware.
These techniques ensure robust and practical neural network deployment on RRAM. 

Although these fault-tolerance techniques enable robust neural network deployment on RRAM-based architectures in practical scenarios, the hardware variations are inevitable and still exist in these hardware platforms, making the RRAM-based NCS vulnerable to malicious attacks in safety-sensitive scenarios.

\subsection{Security Concerns}
As NN-based solutions becoming popular for many applications, NN security arises as a major concern for the practical deployment of NNs in safety-critical tasks.
Lots of efforts have been devoted to investigating the security of NNs, especially, the malicious attack approaches~(\cite{madry2017towards, liu2017fault, zhao2019fault}).
The attack object of these attacks is either the input sample (\ie, adversarial attack) or the neural network weights (\ie, fault injection attack).
The adversarial attack generates adversarial examples by adding invisible perturbations on the input images to fool the neural network, and the fault injection attack poisons the neural network weights via fault injection ways.
In addition, both categories of attack methods are purely software-level attacks and are white-box attacks, that is, the attacker is assumed to have the full knowledge of network model (\eg, model architecture and model parameters, and the information of the hardware platforms to inject faults). 

In general, a successful attack shall satisfy the three criteria: \textit{efficiency, effectiveness} and \textit{stealthiness}. 

\textit{Efficiency}: the attack shall be easy to launch without introducing too much overhead during the attack process. The adversarial attack is naturally efficient than the fault injection attack since it is easy to add perturbations on the input sample. 

\textit{Effectiveness}: the attack shall effectively fool the neural network and also be hard to resist. As the fault injection attack directly distorts the model parameters, it can effectively manipulate the prediction results of neural networks. 

\textit{Stealthiness}: the attacker shall maintain the model accuracy for regular images other than the targeted one. For the adversarial attack, the perturbations added on the input should be human-imperceptible to ensure the attack stealthiness. The fault injection attack should ensure the model accuracy after fault injection.

The adversaries have various demands on these metrics based on the attack scenarios. For instance, the less experienced adversaries can adopt the adversarial attacks, since it is easy to access and perturb the input samples with sacrificed effectiveness. While the adversaries with expert knowledge concern the effectiveness and stealthiness of the attack, they can manipulate the network prediction for a given image at will through fault injection techniques while ensuring model accuracy on other images. Previous studies have demonstrated that it is practical to launch precise and effective fault injection attacks on neural networks \cite{liu2017fault, zhao2019fault}.

In summary, the adversarial attacks focus on the attack stealthiness and  efficiency and the fault injection attacks pursue the attack stealthiness and effectiveness. 
Here, we briefly introduce the representative works on the adversarial attack and fault injection attack, respectively.

\subsubsection{Adversarial Attack} 
There is now a sizable body of works proposing various adversarial attacks which explore different methods to generate effective perturbations~\cite{goodfellow2014explaining, madry2017towards, szegedy2013intriguing}. 
The most representative and effective adversarial attack methods are gradient-based, such as PGD~(\cite{madry2017towards}), FGSM~(\cite{goodfellow2014explaining}), and C\&W~(\cite{carlini2017towards}). 
These attack methods utilize the network gradients with respect to the input to craft the perturbations. For example, the PGD attack iteratively updates the given image in the gradient ascent direction to craft the adversarial examples, and make use of $l_{1}$ norm to measure the image distortion.

\subsubsection{Fault Injection Attack} 
Previous works have investigated different ways to locate and perturb the victim parameters to ensure the stealthiness and efficiency of the fault injection attack~\cite{liu2017fault, zhao2019fault}. 
Single Bias Attack (SBA) and Gradient Descent Attack (GDA) are proposed for different attack scenarios and objectives in ~\cite{liu2017fault}. The SBA only modifies one neuron bias, since the output is linearly dependent on the bias term, to achieve high attack efficiency with relaxed attack stealthiness. 
While GDA applies the gradient descent mechanism to modify the layer-wise parameters. In contrast to SBA, the GDA focus on attack stealthiness instead of attack efficiency. 
The fault sneaking attack (\cite{zhao2019fault}) formulates the fault injection attack as an optimization problem with the constraint of accuracy degradation and parameter modifications, and applies ADMM (alternating direction method of multipliers) to obtain an analytical solution for the parameter modification. 
However, there are still a large number of parameter modifications in the fault sneaking attack, which leads to low attack efficiency.
How to achieve high stealthiness and efficiency at the same time and realize a practical fault injection attack is the main pursuit of this work.

In this work, we incorporate the hardware information, \ie, intrinsic RRAM variations, to improve the performance (\ie, effectiveness, stealthiness and efficiency) of the conventional adversarial attack and fault injection attack.

\section{VADER:Variation-oriented adversarial attack}
In this section, we introduce our VADER whose principle is exploiting the RRAM variation to launch a powerful adversarial attack, such that VADER can even penetrate the adversarial defense, \ie, adversarial training~(\cite{goodfellow2014explaining}). At the same time, the attack stealthiness is ensured.

\subsection{Overview}
\begin{figure*}[t]
  \centering
  \includegraphics[width=\linewidth]{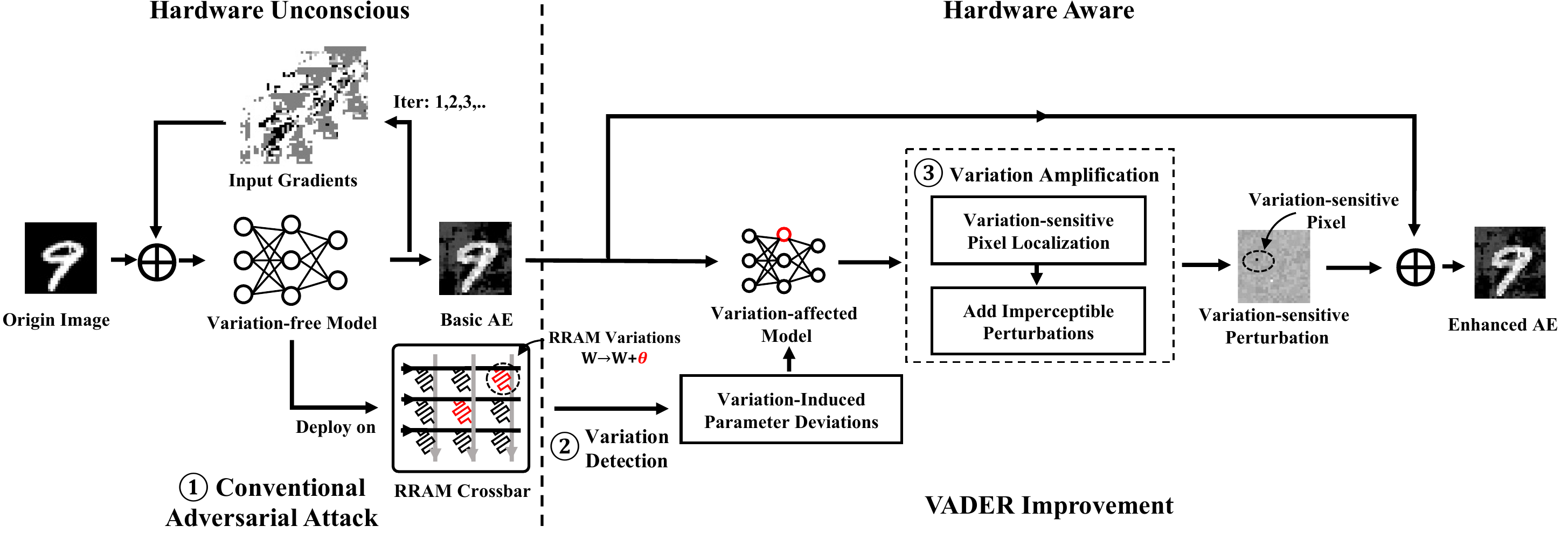}
  \caption{The Workflow of VADER. AE: Adversarial Example.}
  \label{VADER_overview}
  \vspace{\baselineskip}
\end{figure*}
The motivation behind our VADER is that the RRAM variation poses a stealthy security risk because of the variation-induced parameter deviations and the resulting subtle computing errors.
VADER leverages the RRAM variation to aggravate the computing errors from the RRAM-based NCS by adding variation-sensitive perturbation on the input sample. 
The deviation of the computing result will accumulate layer by layer and consequently cause an erroneous prediction for the perturbed input.

\textit{Attack Objective:} the objective of VADER is to find a valid variation-sensitive perturbation for the given input sample to construct the enhanced adversarial example, such that the negative effect of the intrinsic and slight RRAM variation (\eg, parameter deviations and computing errors) can be amplified to disturb the network prediction result. 
Effectiveness and stealthiness are the main optimized goal of our VADER. 

\textbf{Notations.} Before presenting our proposed attack methods, we summarize the notations used in this article in Table~\ref{tab:notations}. The well-trained model is referred to as variation-free before being deployed on RRAM and variation-affected after being deployed on RRAM, respectively. The i-th layer's parameters of the variation-free model are denoted as $W_i$, and $\theta_i$ represents the variation-induced parameter deviations in the i-th layer from the desired value ($W_i$) after mapping the variation-free model on RRAM computing systems.

\begin{table}
  \centering
  \caption{Notations used in this paper.}
  \label{tab:notations}
  \begin{tabularx}{0.45\textwidth}{l|X}
    \toprule
    Notations & Descriptions \\
    \midrule
    $W_i$ & The i-th layer's parameters of the variation-free model, $i=1, 2..$. \\
    \midrule
    $\theta_i$ & The variation-induced parameter deviations in i-th layer, $i=1, 2..$. \\
    \midrule
    $x$ & The input sample. \\
    \midrule
    $y_i(')$ & The i-th layer's output of the variation-free (variation-affected) model, $i=1, 2..$.\\
    \midrule
    $f$  & The Activation functions.\\
    \midrule
    $\mathcal{L}$ & Loss function of the model.\\
    \midrule
    $g(')$ & The gradients of variation-free (variation-affected) model with respect to the input. \\
    \midrule
    $g_d$ & The gradients difference between the variation-free model and  variation-affected model. \\
  \bottomrule
\end{tabularx}
\end{table}

The overall workflow of VADER is illustrated in Fig.~\ref{VADER_overview}.
Given an input sample and a protected network model by the state-of-the-art defense mechanism (\ie, adversarial training), we first perform a \textit{conventional adversarial attack} method (\eg, PGD~\cite{madry2017towards}) to obtain the \textit{basic adversarial example} (\circled 1).
To be specific, we feed the specified sample into the model to compute the gradient, and add the gradient on the origin image. By repeating this operation several times, the basic adversarial example is obtained. 
The basic adversarial example is currently insufficient to deceive the protected network, so we take the following actions to enhance the adversarial example by exploiting the RRAM variation.
Next, we can obtain the RRAM variation-induced parameter deviations (\ie, location of variation-affected parameters and their corresponding values, $W\rightarrow W+\theta$) through the \textit{variation detection} step ~\cite{kannan2015modeling} (\circled 2), which utilizes the inherent sneak-paths to detect the faults and variations in crossbar memories.
The testing method provides high fault coverage, and can locate the exact location of the abnormal RRAM cells.
With the parameter deviations, we can derive the practical variation-affected model, and perform the \textit{variation amplification} (detailed in Section~\ref{var_amp}) on it to generate the \textit{variation-sensitive perturbation} that can amplify the negative impact of the deviated network parameters on the computing results (\circled 3).
Finally, we generated the enhanced adversarial example by adding the \textit{basic adversarial example} with the \textit{variation-sensitive perturbation}.

By being aware of the RRAM variation, the VADER can effectively disable the adversarial defense and mislead the network to produce an erroneous prediction.

\subsection{Variation Amplification}\label{var_amp}
The \textit{variation amplification} stage locates the variation-sensitive pixels in the input image that can amplify the negative effects induced by the RRAM variation, and then determines the perturbation magnitude for these located variation-sensitive pixels.

The variation-sensitive perturbation generation procedure is described in Algorithm~\ref{alg_VADER}.
The inputs of the algorithm include the basic adversarial example $x_{adv}$ from \textit{conventional adversarial attack}, the network classifier $\hat{C}$ that is deployed on the RRAM-based NCS, the learning rate $lr$, and the maximum number of pixels $N_p$ allowed to be perturbed. 
The localization of the variation-sensitive pixels and the decision of their perturbation magnitude is implemented as the nested loop (Line 3-13). 
The outer loop firstly computes $g_d$ (Line 4), then locates the most variation-sensitive pixels by selecting the pixel with the largest $g_d$ (Line 5), and add it into the set of candidate variation-sensitive pixels $S_{p}$ (Line 6). 
Here, the $g_d$ refers to the gradient difference between the variation-affected model (\ie, the network model on RRAM) and the variation-free model (\ie, the network model on GPU).
The inner loop (Line 8-12) decides the value of these candidate pixels through a gradient ascent approach.
Once the number of pixels in $S_p$ reaches $N_p$ or the prediction of the network is misled (Line 13), the Algorithm \ref{alg_VADER} will terminate, and the variation-sensitive perturbation ($\sigma_{vsp}$) is generated. 
The defined $N_p$ is to limit the proportion of the perturbed pixels to the total number of input pixels such that the perturbation will not cause visible attention and thus ensure stealthiness.

\LinesNumbered\IncMargin{1em}
\begin{algorithm}[t]
  \SetKwInOut{Input}{input}\SetKwInOut{Output}{output}
  \Input{The basic adversarial example, $x_{adv}$ ; \\
        Variation affected network, $\hat{C}(W+\theta,\cdot)$; \\
        Learning rate, $lr$; \\
        Maximum number of pixels allowed be \\ perturbed, $N_p$.}
  \Output{Variation-sensitive Perturbation $\delta_{vsp}$.}
  \BlankLine
  \emph{$\delta_{vsp} \leftarrow$ zero matrix of the same shape as $x_{adv}$ \\ $S_p \leftarrow \{\}$ \tcp*[f]{$S_p$ is the coordinate set of variation-sensitive pixels.}} \ 
  \Repeat {(($N_p > 0$) and (the classifier $\hat{C}$ is mislead by $x_{adv}+\delta_{vsp}$))}{
    Compute the gradients $g_d$ of input $x_{adv}+\delta_{vsp}$; \\
    Select the pixel on the input with largest $g_d(i,j)$ as the variation-sensitive pixel; \tcp*[f]{(i,j) is the pixel coordinate;}\\
    $N_p=N_p - 1$;\\
    Insert $(i,j)$ into $S_p$;  \\

    \Repeat {((loss of \ $\hat{C}$ converges) or (the classifier $\hat{C}$ is mislead by $x_{adv}+\delta_{vsp}$))}{
        $\delta_{vsp} \pluseq lr*sign(g_d)$;       \\
        $\delta_{vsp}(i,j) \leftarrow 0,  for\ (i,j)\notin S_p$; \\
        Compute the gradients $g_d$ of input $x_{adv}+\delta_{vsp}$; \\
    }
    }
\caption{Variation-sensitive Perturbation Generation}\label{alg_VADER}
\end{algorithm}\DecMargin{1em}\par

In the following, we will show the effectiveness of $g_d$ in indicating the variation-sensitive pixels of the input image with a running example.

\textbf{Running Example.}
We take a two-layer fully connected network as an example to theoretically demonstrate the feasibility of the gradient difference $g_d$ for locating the variation-sensitive pixels. 
The notations used in the following formulas can refer to Table~\ref{tab:notations}.
The feedforward flow of the network can be formulated as:
\begin{equation}
  y_1 = W_1 * x,
\end{equation}
\begin{equation}
  y_2 = W_2 * f(y_1),
\end{equation}
\begin{equation}
  loss = \mathcal{L}(y_2).
\end{equation}

The gradients $(g$ and $g')$ with respect to the input $x$ can be calculated as:
\begin{equation}
  g = \frac{\partial{\mathcal{L}(y_2)}}{\partial{y_2}}\times W_2^T\odot\frac{\partial{f(y_1)}}{y_1}\times W_1^T,
\end{equation}
\begin{equation}
  g' = \frac{\partial{\mathcal{L}(y_2')}}{\partial{y_2'}}\times (W_2+\theta_2)^T\odot\frac{\partial{f(y_1')}}{y_1'}\times (W_1+\theta_1)^T.
\end{equation}

Theoretically, the gradients of the input pixels with respect to the loss function quantitatively reflect the contribution of each input pixel on the loss value and the prediction result.
On the other hand, since only few parameters are variation-affected, the input gradients of the variation-affected model $g'$ are almost identical to $g$, to be specific, the parameter deviations $\theta_i$ are slight compared to the network parameters $W_i$, and thus the computation result of gradients is dominated by $W_i$, and
therefore it is insufficient to distinguish the variation-sensitive pixels from the less sensitive pixels in the input sample by directly using $g'$. 
To locate the variation-sensitive pixels and amplify the impact of RRAM variation, we subtract the $g$ from $g'$ and the gradient difference $g_d$ is obtained as Equation~\ref{equ:ga}.
\begin{figure*}
\centering
\begin{equation}
  g_d = g'-g
  =\frac{\partial{\mathcal{L}(y_2')}}{\partial{y_2'}}\times (W_2+\theta_2)^T\odot\frac{\partial{f(y_1')}}{y_1'}\times (W_1+\theta_1)^T -
  \frac{\partial{\mathcal{L}(y_2)}}{\partial{y_2}}\times W_2^T\odot\frac{\partial{f(y_1)}}{y_1}\times W_1^T.
\label{equ:ga}
\end{equation}
\end{figure*}

\begin{figure*}
\centering
\begin{equation}
  g_d
\approx \frac{\partial{\mathcal{L}(y_2')}}{\partial{y_2'}}\times \theta_2^T\odot \frac{\partial{f(y_1')}}{y_1'}\times W_1^T + \frac{\partial{\mathcal{L}(y_2')}}{\partial{y_2'}}\times W_2^T \odot\frac{\partial{f(y_1')}}{y_1'}\times\theta_1^T +  \frac{\partial{\mathcal{L}(y_2')}}{\partial{y_2'}}\times \theta_2^T\odot\frac{\partial{f(y_1')}}{y_1'}\times \theta_1^T.
\label{equ:ga_approx}
\end{equation}
\end{figure*}

The outputs of the nonlinear activation layers ($y_i$ and $y_i'$) are nearly identical and that their derivatives ($\frac{\partial{\mathcal{L}(y_i')}}{\partial{y_i'}}$ and $\frac{\partial{\mathcal{L}(y_i)}}{\partial{y_i}}$ ) can be viewed as being the same scalar tensor because the variation-resilient training strategy \cite{xia2017fault} is adopted to mitigate the negative impact of RRAM variations on neural networks (detailed in \ref{exp_setup}). Since Equation~\ref{equ:ga} only comprises the distributive operations of matrix multiplication and Hadamard product, it can be approximated and simplified as Equation~\ref{equ:ga_approx}.
Note that each term in Equation~\ref{equ:ga_approx} is scaled by the parameter deviations $\theta_i$. As such, the gradient difference $g_d$ not only reflects the impact of the input pixels on model predictions, but also indicates their variation sensitivity. To this end, the pixel with the largest $g_d$ has the strongest capability to activate the negative effects of the RRAM variation, and distort the network predictions.

\section{EFI: Enhanced Fault Injection Attack}

\subsection{Overview}
Considering the intrinsic RRAM variation, we develop a lightweight fault injection attack, EFI, which significantly reduces the cost for the engineer-expensive fault injection operation.

\textit{Attack Objective:}  The EFI, as the fault injection attack, primarily targets achieving high attack efficiency and stealthiness.
For the attack efficiency, the EFI aims to minimize required network parameter modifications for a successful attack by leveraging the information of the RRAM intrinsic variations, and thus save the fault injection cost.
As for the attack stealthiness, the EFI shall make the network model predict a specified label with the target input sample while outputting the correct prediction for other input samples.

\begin{figure*}[t]
  \centering
  \includegraphics[width=\linewidth]{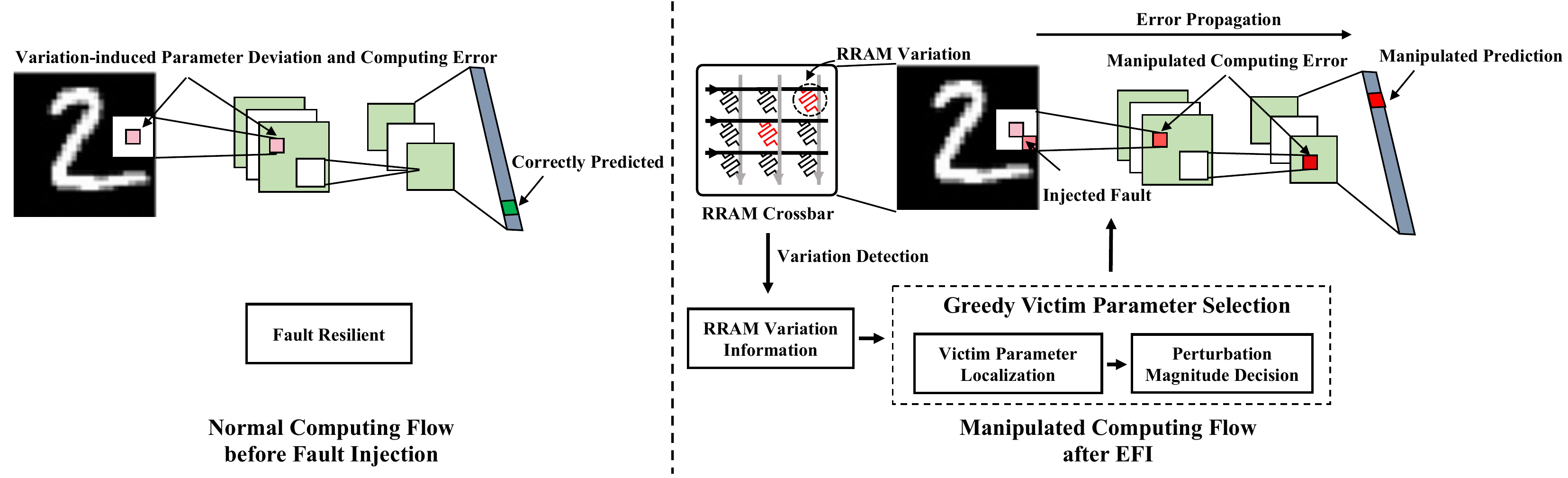}
  \caption{The Workflow of EFI. }
  \label{EFI_overview}
\end{figure*}

The attack flow of EFI is illustrated in Fig~\ref{EFI_overview}.
Normally, the subtle variation-induced parameter and computing error can be tolerated by the error-resilience of neural networks and mitigated by the fault-resilient training strategy, and thus maintaining prediction accuracy.
In the EFI, we inject faults into the target layers to incorporate with these existing RRAM variations to tamper the output feature map, and the distorted output will propagate forward and cooperate with the intrinsic RRAM variations and injected faults in the following layers to manipulate the prediction result.
The overall workflow of EFI is similar to the attack paradigm of VADER. Firstly, we acquire the information of variation-affected parameters through a \textit{variation detection} process. 
Next, we perform \textit{greedy victim parameter selection} (detailed in Section~\ref{greedy_vps}) to select and perturb the desired victim parameters. 
With these perturbed parameters, we can easily manipulate the computing flow of the variation-affected model, and finally change the prediction result to the target label.

\subsection{Greedy Victim Parameter Selection}\label{greedy_vps}
\LinesNumbered\IncMargin{1em}
\begin{algorithm}[t]
  \SetKwInOut{Input}{input}\SetKwInOut{Output}{output}\SetKwFunction{Pos}{Pos\_of}\SetKwFunction{Max}{Max}
  \Input{Specified input sample, $x_s$;\\
        Target label, $y_t$\\
        Variation-affected model, $\hat{C}(W+\theta,\cdot)$;\\
        List of target layers to inject faults, $L$ \\
        Learning rate of each target layers, $lr$.}
  \Output{The set of victim parameters in the target layers, $vp$.}
  \BlankLine
  \For{$vp_i \in vp$ } {
    $vp_i \leftarrow \{\}$ ;
  }
  \Repeat {($\hat{C}(W+\theta,x_s) = y_t$)}{
    \For{$L_i \in L$} 
    {
        Compute the gradients $G_i$ of the target layers $L_i$ with input $x_s$; \\
        Select the parameter $p$ with the largest gradient in $G_i$ as victim parameter;\\
        Insert the victim parameter $p$ into $vp_i$. \\
    } 
    \For{$vp_i \in vp$ } {
        update parameters in $vp_{i}$ in the gradient ascent direction with learning rate $lr_i$ until the loss of $\hat{C}$ converges; \\
    }
  }
\caption{Greedy Victim Parameter Selection}\label{alg_EFI}
\end{algorithm}\DecMargin{1em}\par

The \textit{greedy victim parameter selection} realizes the objective that poisoning the minimal number of victim parameters by incorporating with the existed parameter deviation induced by the RRAM variations.
Algorithm~\ref{alg_EFI} presents the detailed flow of the victim parameter selection. 
The inputs of the algorithm include the specified input sample $x_s$, the target label $y_t$, the variation-affected network classifier $\hat{C}$, the list of target layers $L$ to inject fault and the learning rate $lr$ for these layers.
The selection and perturbation for the victim parameters are also realized by two loops (Line 5-12).
These two loops realize victim parameter localization and perturbation magnitude decision respectively.
In the first loop (Line 5-9), we first compute the gradient $G_i$ for each target layer $L_i$ (Line 6), then, we apply the greedy strategy that selects the parameter with the largest gradient value in $G_i$ as the victim parameter for each target layer (Line~7).
In the second loop (line 10-12), we decide the perturbation magnitude of these victim parameters with gradient ascent.
These two loops will be iterated until the specified input $x_s$ is recognized as the target label $y_t$ (Line 13).
Note that, the learning rate $lr_i$ for different target layers should be carefully adjusted to avoid oversized modifications on the victim parameters and cause significant degradation in the model performance.

\section{Evaluations}

\subsection{Experimental Setup}\label{exp_setup}

\textbf{Datasets and Network Architectures.}
We evaluate VADER and EFI with two different sizes of networks, i.e., LeNet~(\cite{lecun1998gradient}) and Wide Residual Networks (WRN) (\cite{zagoruyko2016wide}) on two classification datasets, MNIST and Cifar10, and compare them with the SOTA attack approaches (\cite{madry2017towards, zhao2019fault, liu2017fault}) in terms of attack effectiveness, efficiency and stealthiness. 
The MNIST is a handwriting digit classification dataset, and Cifar10 is an RGB image classification dataset. 
Both datasets consist of 10 exclusive categories. 
In the experiments, we adopt the variation-resilient training method to ensure the model robustness on RRAM, and both networks are protected with the adversarial training (\cite{goodfellow2014explaining}) against the adversarial attack.

\textbf{Configurations.}
To precisely simulate the impact of RRAM variation on the neural networks, we modify the TensorFlow framework to model the network inference flow on the RRAM-based NCS, and the RRAM variation modeling (\eg, variation distribution and their magnitude) is referred from \cite{he2019noise}.
The evaluated models are quantized to 8-bit precision such that can be deployed on RRAM-based NCS. 
The hyperparameter $N_p$ in Algorithm~\ref{alg_VADER} is set to 10.
The performance metrics are averaged over 10000 random sampled images from the evaluation dataset. 

\subsection{Security Risk of Hardware Variations}\label{security_risk}

\begin{table}
  \caption{Performance Comparison of Models on Different Hardware Platforms. The Acc and Acc* denote the network accuracy on the testing examples and adversarial examples, respectively.}
  \label{tab:table_performance}
  \center
  \begin{tabular}{ccccc}
    \toprule
    Dataset &  \makecell{Network \\Architecture}& Hardware Platform & Acc & Acc* \\
    \midrule
    \multirow{2}{*}{MNIST} & \multirow{2}{*}{LeNet} & GPU & 98.70\% & 88.85\% \\
    ~ & ~& RRAM & 98.83\% & 87.67\%\\
    \cmidrule{1-5}
    \multirow{2}{*}{Cifar10} & \multirow{2}{*}{WRN} & GPU & 85.11\% & 49.68\%  \\
    ~ & ~ & RRAM & 91.13\% & 30.90\% \\
  \bottomrule
  \vspace{\baselineskip}
\end{tabular}
\end{table}
In this section, we investigate the potential security risk of RRAM variations for the neural networks. 
Specifically, we evaluate the impact of RRAM variation on the network adversarial robustness and recognition capacity by comparing the accuracy of variation-affected model and variation-free model on the clean testing images and adversarial examples.
As Table~\ref{tab:table_performance} shows, with the variation-resilient training, the accuracy of the variation-affected models is comparable to, and even slightly higher than, the variation-free models, while there is a degradation in the resistance of the adversarial examples for the variation-affected model (1.08\% and 18.78\% on the MNIST and Cifar10, respectively).
The results indicate that although the RRAM variations are harmless to the practical application of neural networks, they indeed impair the robustness of the model and pose a potential security threat to the neural network.

\subsection{Performance Evaluation}
In the following paragraphs, we will systematically evaluate the effectiveness, efficiency and stealthiness for our proposed VADER and EFI.

\subsubsection{Evaluation for VADER}
In this section, we evaluated the attack effectiveness and stealthiness of VADER on the MNIST and Cifar10 datasets and a variety of RRAM devices.
Specifically, we use the attack success rate to measure attack effectiveness, and is defined as:
\begin{equation}
  \textnormal{Attack Success Rate (ASR)} = \frac{\textnormal{\# of mis-classified samples}}{\textnormal{\# of evaluated samples}}.
\end{equation}

\begin{figure}[t]
  \centering
  \includegraphics[width=0.95\linewidth]{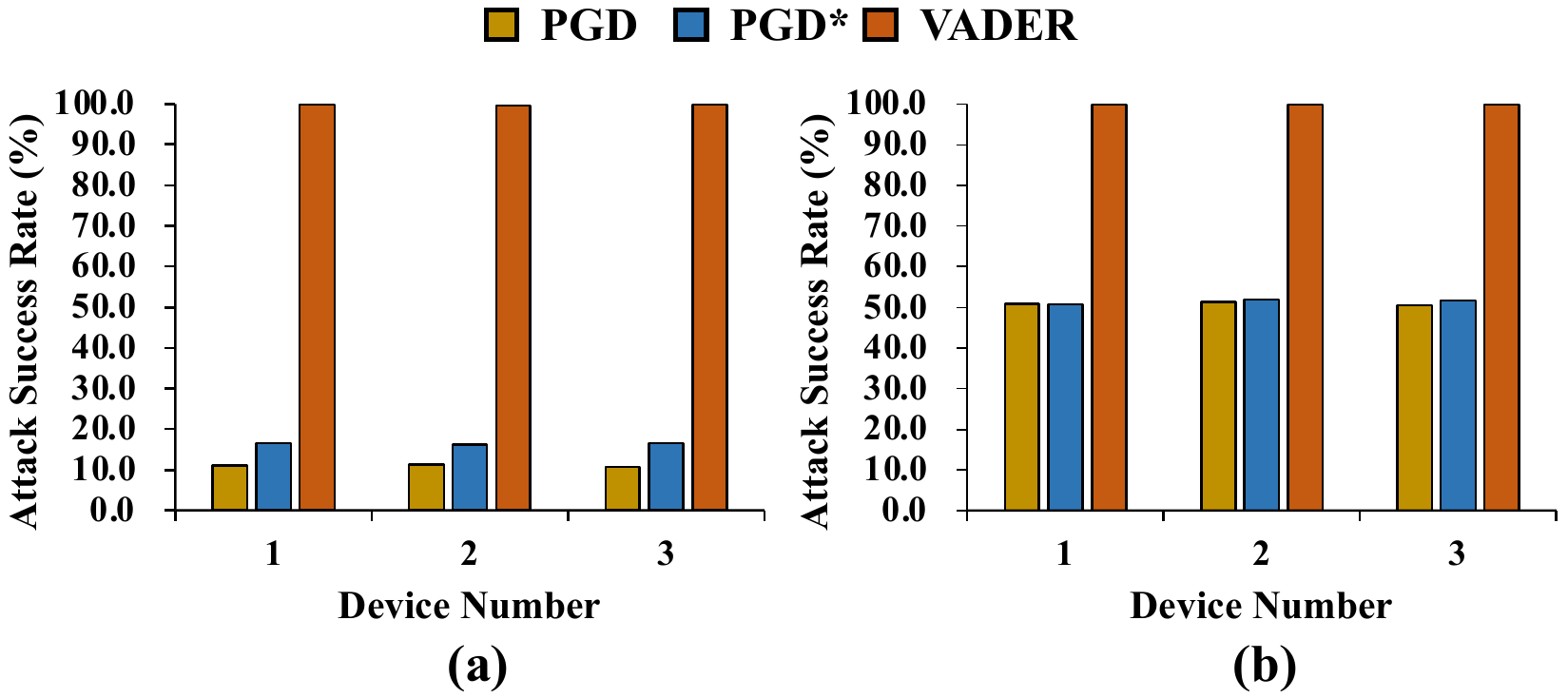}
  \caption{Effectiveness evaluation of VADER. (a) and (b) are evaluated on MNIST and Cifar10 respectively.}\label{vader_effectiveness}
\end{figure}

As results shown in Figure~\ref{vader_effectiveness}, conventional adversarial attack, i.e. PGD is resisted by the adversarial defense mechanism, while VADER can achieve nearly 100\% attack success rate, demonstrating its superior effectiveness.
Furthermore, to analyze the benefit of the hardware knowledge in our VADER, we design an enhanced PGD, denoted as PGD*, which mimic the \textit{variation amplification} step in Algorithm~\ref{alg_VADER} by substituting the gradient difference $g_a$ with the gradient of the variation-free model $g$.
In contrast to VADER, PGD* is unaware of the hardware platforms and RRAM variation. 
As the results show, PGD* achieves slightly higher ASR than PGD, but is still incomparable to our VADER.
Besides, VADER can maintain effective consistency on different RRAM devices with different variation distributions.
These results indicate that the hardware information and localization of the variation-sensitive pixels in \textit{variation amplification} step can improve the attack effectiveness.

\begin{figure}[t]
  \centering
  \includegraphics[width=0.95\linewidth]{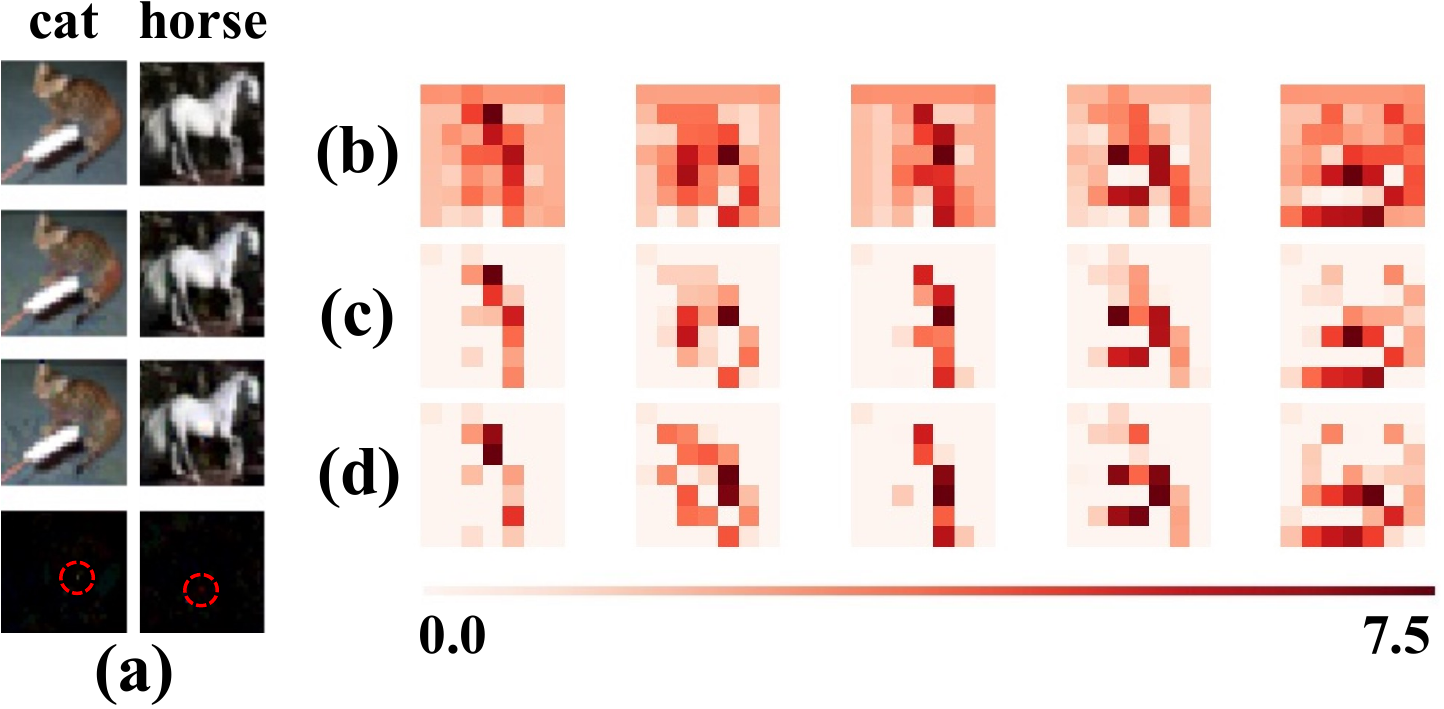}
  \caption{ (a) Visualization of VADER generated enhanced adversarial examples from Cifar10. (b)(c)(d) Visualizing samples of output feature maps from the final pooling layer.}\label{vader_visual}
  \vspace{\baselineskip}
\end{figure}

VADER realizes the objective of stealthy attack by adding human imperceptible perturbations on the input samples. 
Here, we visualize several enhanced adversarial examples from VADER in Figure~\ref{vader_visual}(a). 
From top to bottom, the images in each row are original clean images, adversarial examples from PGD, enhanced adversarial examples from VADER, and the difference between these two adversarial examples, respectively. 
As the figure shows, our VADER adversarial examples are visually indistinguishable from the PGD adversarial examples and even the original clean images, except for the several perturbed pixels. The perturbed pixels are highlighted by the red circles, which are too slight to cause human notice. 
This result guarantees the stealthiness of VADER.

Furthermore, to verify that VADER can satisfy our design principle (i.e. VADER is variation-oriented), we visualize several activation outputs of the network trained on MNIST in Figure~\ref{vader_visual}(b)(c)(d). 
All these feature maps are sampled from the output of the last pooling layers. 
Respectively, Figure~\ref{vader_visual}(b) shows the feature maps from the variation-affected model with VADER adversarial examples as input. Figure~\ref{vader_visual}(c) shows the feature maps from the variation-free model with VADER adversarial examples as input. Figure~\ref{vader_visual}(d) shows the feature maps from the variation-free model with clean test image as input.
The feature maps in the last two rows are highly similar, while the feature maps in the first row are visually different from the other two rows. 
This observation indicates that the VADER can only interfere with the computing flow of the variation-affected model, and thus VADER is variation-oriented. 
Besides, the visual difference suggests that VADER can satisfy our design principle of leveraging the RRAM variation to lead the activation output to deviate from the desired output and propagate the computational error forward. 
In summary, VADER can facilitate the RRAM variation to improve the attack effectiveness and realize a stealthy adversarial attack at the same time.

\subsubsection{Evaluations for EFI}
In this section, we evaluate the efficiency and stealthiness of EFI.
We use the number of the modified network parameters as a proxy measure of the cost of a fault injection attack.
To compute the fault injection cost, we randomly sampled 100 test images and labels, and averaged the number of parameter modifications required for a successful EFI attack.
We compare the EFI with the state-of-the-art fault injection attacks, including Single Bias Attack (SBA), Gradient Descent Attack (GDA) from \cite{liu2017fault} and Fault Sneaking Attack from \cite{zhao2019fault}. The evaluation results are summarized in Table~\ref{tab:EFI_eval}. 
In contrast to the stealthy Fault Sneaking Attack and GDA, EFI can save \textgreater 95\% fault injection operations, while maintaining higher classification accuracy after the fault injection attack. Compared with SBA, EFI improves the accuracy degradation by a significant margin.
These results demonstrate that EFI can significantly improve the attack efficiency and stealthiness, and ensure a well balance between the two metrics. 

To understand the EFI better, we perform EFI on the variation-affected model and the variation-free models to analyze the benefit from the hardware variations.
As seen in Table~\ref{tab:EFI_eval}, EFI can not mislead the variation-free model.
This indicates that our EFI is also a variation-oriented attack method, and can exploit these variations to reduce the engineering cost and improve the stealthy.
Furthermore, we perform victim parameter selection on different layer combinations to investigate the impact of fault-injected layers.
The result shows that the fault injection on the later layers can achieve better stealthiness.
The reason behind this observation may be that the latter network layers are closer to the prediction layer, and have a stronger correlation on the prediction results.

\begin{table}
\centering
  \caption{The Performance Evaluation of EFI on RRAM Platform. The AD and MP are abbreviation for accuracy degradation and modified parameters.}
  \label{tab:EFI_eval}
  \begin{tabular}{c|c|c|c|c}
    \toprule
    \multirow{2}{*}{Attack Method}  & \multicolumn{2}{c|}{MNIST}  & \multicolumn{2}{c}{Cifar10}\\
    \cmidrule{2-5}
    ~ & AD& \# of MP & AD & \# of MP \\
    \midrule
    SBA  & - & - & -24.4\% & 1\\
    GDA & -3.86\% & 1170 & -2.35\% &  198\\
    Fault Sneaking & -0.80\% & \textgreater 1026 & -1.0\% & \textgreater 1026 \\
    \cmidrule{1-5}
    \makecell{EFI \\(on variation-free model)} & \multicolumn{2}{c|}{failed} & \multicolumn{2}{c}{failed} \\
    \cmidrule{1-5}
    EFI (layer 1, 2) & -3.0\% & 160 & -4.1\% & 30 \\
    EFI (layer 3, 4) & -0.50\% & 40 & -4.8\% & 68 \\
    \cmidrule{1-5}
    EFI (last two layers) & -0.50\% & 40 & -0.9\% &20 \\
  \bottomrule
\end{tabular}
\end{table}

\subsection{Discussion}
In this section, we discuss and analyze the characteristic of VADER and EFI, and finally discuss the defense technique against them.
The fundamental improvement of VADER over the conventional adversarial attack methods is that our VADER is hardware aware. 
Therefore, we attribute the success of VADER to the insufficiency of adversarial defense techniques. The adversarial defense techniques are purely software-level, and unaware of the hardware variations.
To protect the network model against our VADER, it is essential for the defense mechanism to be hardware-aware.
Thus, we suggest an on-device adversarial training defense method that protects the model by embedding hardware variation information into the adversarial training process, and experimental result shows the hardware-aware defense can resist VADER.
Regrading the defense for EFI, we advise retaining a fault detection routine as \cite{li2019rramedy}, which periodically identifies potential fault injection attacks using a testing preserved dataset. 
If the tested predicted confidence score significantly deviates from the groundtruth confidence score in the testing set, it is likely that the tested model has suffered from a fault injection attack.
After the fault injection attack is confirmed, we can compare the parameters on the device with the backup parameters and reprogram the different parameters to recover the injected faults.

\section{conclusion}
In this paper, we revealed that the intrinsic RRAM variation poses a security risk for the RRAM-based NCS, and therefore propose two hardware-aware attack methods, VADER and EFI, which leverage the RRAM variation to improve the attack performance metrics, including effectiveness, efficiency and stealthiness.
The experimental results show that both improved attack methods can achieve almost 100\% attack success rate with minimized operational cost, demonstrating superior performance than the conventional algorithmic attack methods.
Besides, the two attack methods are orthogonal and can be combined to achieve better performance.
Finally, we systematically analyze the proposed attack methods and discuss the feasible defense mechanism to eliminate the threat from VADER and EFI.

\section{Acknowledgements}
This paper is supported in part by the National Natural Science Foundation of China (NSFC) under grant No.(61874124, 62204164, 62222411), Zhejiang Lab under Grants 2021PC0AC01.
This paper is also supported by Beijing Natural Science Foundation (4194092).


\begin{thebibliography}{10}
\providecommand{\url}[1]{#1}
\csname url@samestyle\endcsname
\providecommand{\newblock}{\relax}
\providecommand{\bibinfo}[2]{#2}
\providecommand{\BIBentrySTDinterwordspacing}{\spaceskip=0pt\relax}
\providecommand{\BIBentryALTinterwordstretchfactor}{4}
\providecommand{\BIBentryALTinterwordspacing}{\spaceskip=\fontdimen2\font plus
\BIBentryALTinterwordstretchfactor\fontdimen3\font minus
  \fontdimen4\font\relax}
\providecommand{\BIBforeignlanguage}[2]{{%
\expandafter\ifx\csname l@#1\endcsname\relax
\typeout{** WARNING: IEEEtran.bst: No hyphenation pattern has been}%
\typeout{** loaded for the language `#1'. Using the pattern for}%
\typeout{** the default language instead.}%
\else
\language=\csname l@#1\endcsname
\fi
#2}}
\providecommand{\BIBdecl}{\relax}
\BIBdecl

\bibitem{he2016deep}
K.~He, X.~Zhang, S.~Ren, and J.~Sun, ``Deep residual learning for image
  recognition,'' in \emph{Proceedings of the IEEE conference on computer vision
  and pattern recognition}, 2016, pp. 770--778.

\bibitem{simonyan2014very}
K.~Simonyan and A.~Zisserman, ``Very deep convolutional networks for
  large-scale image recognition,'' \emph{arXiv preprint arXiv:1409.1556}, 2014.

\bibitem{ren2015faster}
S.~Ren, K.~He, R.~Girshick, and J.~Sun, ``Faster r-cnn: Towards real-time
  object detection with region proposal networks,'' \emph{Advances in neural
  information processing systems}, vol.~28, pp. 91--99, 2015.

\bibitem{devlin2018bert}
J.~Devlin, M.-W. Chang, K.~Lee, and K.~Toutanova, ``Bert: Pre-training of deep
  bidirectional transformers for language understanding,'' \emph{arXiv preprint
  arXiv:1810.04805}, 2018.

\bibitem{sutskever2014sequence}
I.~Sutskever, O.~Vinyals, and Q.~V. Le, ``Sequence to sequence learning with
  neural networks,'' in \emph{Advances in neural information processing
  systems}, 2014, pp. 3104--3112.

\bibitem{jo2010nanoscale}
S.~H. Jo, T.~Chang, I.~Ebong, B.~B. Bhadviya, P.~Mazumder, and W.~Lu,
  ``Nanoscale memristor device as synapse in neuromorphic systems,'' \emph{Nano
  letters}, vol.~10, no.~4, pp. 1297--1301, 2010.

\bibitem{indiveri2013integration}
G.~Indiveri, B.~Linares-Barranco, R.~Legenstein, G.~Deligeorgis, and
  T.~Prodromakis, ``Integration of nanoscale memristor synapses in neuromorphic
  computing architectures,'' \emph{Nanotechnology}, vol.~24, no.~38, p. 384010,
  2013.

\bibitem{xia2016switched}
L.~Xia, T.~Tang, W.~Huangfu, M.~Cheng, X.~Yin, B.~Li, Y.~Wang, and H.~Yang,
  ``Switched by input: Power efficient structure for rram-based convolutional
  neural network,'' in \emph{2016 53nd ACM/EDAC/IEEE Design Automation
  Conference (DAC)}.\hskip 1em plus 0.5em minus 0.4em\relax IEEE, 2016, pp.
  1--6.

\bibitem{shafiee2016isaac}
A.~Shafiee, A.~Nag, N.~Muralimanohar, R.~Balasubramonian, J.~P. Strachan,
  M.~Hu, R.~S. Williams, and V.~Srikumar, ``Isaac: A convolutional neural
  network accelerator with in-situ analog arithmetic in crossbars,'' \emph{ACM
  SIGARCH Computer Architecture News}, vol.~44, no.~3, pp. 14--26, 2016.

\bibitem{chi2016prime}
P.~Chi, S.~Li, C.~Xu, T.~Zhang, J.~Zhao, Y.~Liu, Y.~Wang, and Y.~Xie, ``Prime:
  A novel processing-in-memory architecture for neural network computation in
  reram-based main memory,'' \emph{ACM SIGARCH Computer Architecture News},
  vol.~44, no.~3, pp. 27--39, 2016.

\bibitem{song2017pipelayer}
L.~Song, X.~Qian, H.~Li, and Y.~Chen, ``Pipelayer: A pipelined reram-based
  accelerator for deep learning,'' in \emph{2017 IEEE International Symposium
  on High Performance Computer Architecture (HPCA)}.\hskip 1em plus 0.5em minus
  0.4em\relax IEEE, 2017, pp. 541--552.

\bibitem{li2020hitm}
B.~Li, Y.~Wang, and Y.~Chen, ``Hitm: high-throughput reram-based pim for
  multi-modal neural networks,'' in \emph{Proceedings of the 39th International
  Conference on Computer-Aided Design}, 2020, pp. 1--7.

\bibitem{li2020red}
Z.~Li, B.~Li, Z.~Fan, and H.~Li, ``Red: A reram-based efficient accelerator for
  deconvolutional computation,'' \emph{IEEE Transactions on Computer-Aided
  Design of Integrated Circuits and Systems}, vol.~39, no.~12, pp. 4736--4747,
  2020.

\bibitem{goodfellow2014explaining}
I.~J. Goodfellow, J.~Shlens, and C.~Szegedy, ``Explaining and harnessing
  adversarial examples,'' \emph{arXiv preprint arXiv:1412.6572}, 2014.

\bibitem{madry2017towards}
A.~Madry, A.~Makelov, L.~Schmidt, D.~Tsipras, and A.~Vladu, ``Towards deep
  learning models resistant to adversarial attacks,'' \emph{arXiv preprint
  arXiv:1706.06083}, 2017.

\bibitem{liu2017fault}
Y.~Liu, L.~Wei, B.~Luo, and Q.~Xu, ``Fault injection attack on deep neural
  network,'' in \emph{2017 IEEE/ACM International Conference on Computer-Aided
  Design (ICCAD)}.\hskip 1em plus 0.5em minus 0.4em\relax IEEE, 2017, pp.
  131--138.

\bibitem{zhao2019fault}
P.~Zhao, S.~Wang, C.~Gongye, Y.~Wang, Y.~Fei, and X.~Lin, ``Fault sneaking
  attack: A stealthy framework for misleading deep neural networks,'' in
  \emph{2019 56th ACM/IEEE Design Automation Conference (DAC)}.\hskip 1em plus
  0.5em minus 0.4em\relax IEEE, 2019, pp. 1--6.

\bibitem{szegedy2013intriguing}
C.~Szegedy, W.~Zaremba, I.~Sutskever, J.~Bruna, D.~Erhan, I.~Goodfellow, and
  R.~Fergus, ``Intriguing properties of neural networks,'' \emph{arXiv preprint
  arXiv:1312.6199}, 2013.

\bibitem{kannan2015modeling}
S.~Kannan, N.~Karimi, R.~Karri, and O.~Sinanoglu, ``Modeling, detection, and
  diagnosis of faults in multilevel memristor memories,'' \emph{IEEE
  Transactions on Computer-Aided Design of Integrated Circuits and Systems},
  vol.~34, no.~5, pp. 822--834, 2015.

\bibitem{he2019noise}
Z.~He, J.~Lin, R.~Ewetz, J.-S. Yuan, and D.~Fan, ``Noise injection adaption:
  End-to-end reram crossbar non-ideal effect adaption for neural network
  mapping,'' in \emph{Proceedings of the 56th Annual Design Automation
  Conference 2019}, 2019, pp. 1--6.

\bibitem{liu2017rescuing}
C.~Liu, M.~Hu, J.~P. Strachan, and H.~Li, ``Rescuing memristor-based
  neuromorphic design with high defects,'' in \emph{2017 54th ACM/EDAC/IEEE
  Design Automation Conference (DAC)}.\hskip 1em plus 0.5em minus 0.4em\relax
  IEEE, 2017, pp. 1--6.

\bibitem{li2019build}
B.~Li, B.~Yan, C.~Liu, and H.~Li, ``Build reliable and efficient neuromorphic
  design with memristor technology,'' in \emph{Proceedings of the 24th Asia and
  South Pacific Design Automation Conference}, 2019, pp. 224--229.

\bibitem{hu2016dot}
M.~Hu, J.~P. Strachan, Z.~Li, E.~M. Grafals, N.~Davila, C.~Graves, S.~Lam,
  N.~Ge, J.~J. Yang, and R.~S. Williams, ``Dot-product engine for neuromorphic
  computing: Programming 1t1m crossbar to accelerate matrix-vector
  multiplication,'' in \emph{2016 53nd ACM/EDAC/IEEE Design Automation
  Conference (DAC)}.\hskip 1em plus 0.5em minus 0.4em\relax IEEE, 2016, pp.
  1--6.

\bibitem{xia2017fault}
L.~Xia, M.~Liu, X.~Ning, K.~Chakrabarty, and Y.~Wang, ``Fault-tolerant training
  with on-line fault detection for rram-based neural computing systems,'' in
  \emph{Proceedings of the 54th Annual Design Automation Conference 2017},
  2017, pp. 1--6.

\bibitem{carlini2017towards}
N.~Carlini and D.~Wagner, ``Towards evaluating the robustness of neural
  networks,'' in \emph{2017 ieee symposium on security and privacy (sp)}.\hskip
  1em plus 0.5em minus 0.4em\relax IEEE, 2017, pp. 39--57.

\bibitem{lecun1998gradient}
Y.~LeCun, L.~Bottou, Y.~Bengio, and P.~Haffner, ``Gradient-based learning
  applied to document recognition,'' \emph{Proceedings of the IEEE}, vol.~86,
  no.~11, pp. 2278--2324, 1998.

\bibitem{zagoruyko2016wide}
S.~Zagoruyko and N.~Komodakis, ``Wide residual networks,'' \emph{arXiv preprint
  arXiv:1605.07146}, 2016.

\bibitem{li2019rramedy}
W.~Li, Y.~Wang, H.~Li, and X.~Li, ``Rramedy: Protecting reram-based neural
  network from permanent and soft faults during its lifetime,'' in \emph{2019
  IEEE 37th International Conference on Computer Design (ICCD)}.\hskip 1em plus
  0.5em minus 0.4em\relax IEEE, 2019, pp. 91--99.

\end{thebibliography}
\end{document}